\documentclass[twocolumn,preprintnumbers,amsmath,amssymb]{revtex4}
\usepackage{dcolumn}
\usepackage{bm}
\usepackage{graphicx}
\usepackage{amsmath}

\begin{document}

\title{Exact solutions for a spin-orbit coupled bosonic double-well system}
\author{Yunrong Luo$^{1}$\footnote{Corresponding author: lyr\underline{ }1982@hunnu.edu.cn}, \ Xuemei Wang$^{1}$, \ Jia Yi$^{2}$, \ Wenjuan Li$^{1}$, \ Xin Xie$^{1}$, and \ Wenhua Hai$^{1}$}
\affiliation{$^{1}$Key Laboratory of Low-dimensional Quantum Structures and Quantum Control of Ministry
of Education, and Key Laboratory for Matter Microstructure and Function of Hunan Province, School of Physics and Electronics, Hunan Normal University, Changsha 410081, China\\
$^{2}$School of Management, Hunan University of Information Technology, Changsha 410151, China}

\begin{abstract}

Exact solutions for spin-orbit (SO) coupled cold atomic systems are very important and rare in physics. In this paper, we propose a simple method of combined modulations to generate the analytic exact solutions for an SO-coupled boson held in a driven double well. For the cases of synchronous combined modulations and the spin-conserving tunneling, we obtain the general analytical accurate solutions of the system respectively. For the
case of spin-flipping tunneling under asynchronous combined modulations, we get the special exact solutions in simple form when the driving parameters satisfy certain conditions.
Based on these obtained exact solutions, we reveal some intriguing quantum spin dynamical phenomena, for instance, the arbitrary population transfer (APT) with and/or without spin-flipping, the controlled coherent population conservation (CCPC), and the controlled coherent population inversion (CCPI). The results may have potential applications in the preparation of accurate quantum entangled states and quantum information processing.

\end{abstract}

\maketitle

\section{Introduction}

Exact solutions have played a central role in many branches of physics\cite{barnes109}, due to the fact that they not only can supplement numerical simulations, but also can deepen understanding of the underlying physics\cite{Zakrzewski32}. Therefore, since the birth of quantum mechanics, the search for exact solutions of quantum systems has been ongoing\cite{barnes109}. Although widely studied over the past few decades, up to now many accurate solutions are obtained, which are based on two-level (or two-state) systems\cite{barnes109, Zakrzewski32, landau2, zener137, rosen40, rabi51, hioe32, hioe30, hai87, luoxb95, xie82, vitanov9, JC51}, such as the paradigmatic Landau-Zener\cite{landau2, zener137} and Rabi\cite{rabi51}models, driven two-level systems\cite{barnes109, xie82}, complete population inversion by a phase jump for a two-state system\cite{vitanov9}, the  Jaynes-Cummings model\cite{JC51}, and so on. The accurate analytic solution has proved to be very important in the contexts of qubit control\cite{gre5, poem107, hai29} and has been extended to a series of precise controls\cite{Zakrzewski32, hioe30, hai87, luoxb95, xie82, vitanov9, bambini23, kyo71, gang82}. Not only that, the accurate solutions are extremely useful for the fundamental importance of quantum systems.
However, for quantum systems with more levels, the exact solutions are more difficult to handle and are still lacking\cite{bere, llorente}.

In recent years, the study of spin-orbit (SO) coupling has been a hot topic, which has led to a number of interesting physical phenomena\cite{kato306, bernevig314, zutic76, step93}. It is found that SO coupling is specially appropriate for qubit control in the spintronic devices\cite{zutic76, step93, flindt97, bednarek101} and plays a crucial role in topological quantum computing\cite{nayak80}. Synthetic SO coupling of both Bose and Fermi gases has been realized in recent experiments\cite{lin471, wang109, cheuk109, zhang109, huang12, wu354, zhang128}, which opens a completely new and ideal avenue for investigating quantum dynamics of SO-coupled cold atomic systems. There have been a great number of research works focusing on novel dynamics of SO-coupled cold atomic systems\cite{kart117, zhang628, tang121, wu99, ng92, zhang609, garcia89, citro224, tang, cheng89, qu19, orso118, chen86, hu93, olson90, zcw108, yu90, ji99, luo93, luo52, luo22, gar114, abdu98, luo2021, luo39, luo74}, for instance, quantum spin dynamics\cite{tang121, wu99}, spin Josephson dynamics\cite{zhang609, garcia89, citro224, tang} and localization\cite{cheng89, qu19}, Anderson transition of SO-coupled cold atoms\cite{orso118}, collective dynamics\cite{zhang109, hu93}, Bloch oscillation dynamics\cite{ji99}, selective coherent spin transportation\cite{yu90}, spin-dependent dynamic localization\cite{luo93}, controlling stable spin dynamics\cite{luo22}, transparent manipulation of spin dynamics\cite{luo39}, coherent control of spin tunneling\cite{luo74}, and so on. To our knowledge, many works related to the quantum dynamics of SO-coupled cold atomic systems have been studied by applying approximate methods\cite{hu93, tang, olson90, zcw108, ji99, cheng89, qu19, orso118, chen86, yu90, luo93, luo22, citro224, gar114, abdu98, luo52, luo2021}, for instance, numerical simulation\cite{hu93, olson90, zcw108, ji99, cheng89, qu19, orso118}, variational approximation\cite{hu93, cheng89,chen86}, mean-field approximation\cite{zcw108}, high-frequency approximation(or rotating-wave approximation)\cite{yu90, luo93, luo22, citro224, tang, gar114, abdu98, luo39, luo74}, multiple-time-scale asymptotic analysis\cite{luo52, luo2021}, etc. However, the research on quantum spin dynamics of SO-coupled ultracold atomic systems based on exact solutions is still extremely rare\cite{hai29}.

In this paper, we present a simple method to construct the exact analytic solutions by applying combined modulations for an
SO-coupled boson trapped in a double well potential (a four-level or four-state system). For the synchronous combined modulations case and the spin-conserving tunneling case, we obtain the analytical accurate solutions in general forms, respectively. For the spin-flipping case under asynchronous combined modulations, when the driving parameters are appropriately fixed, we get the simple particular exact solutions.
Based on the obtained analytical exact solutions, some interesting quantum spin dynamical phenomena are revealed. Under synchronous combined modulations, the arbitrary population transfer (APT) with and/or without spin-flipping, the controlled coherent population conservation (CCPC), and the controlled coherent population inversion (CCPI) between left and right wells are shown. Under asynchronous combined modulations, for the spin-conserving and spin-flipping cases, we find new parametric conditions for performing the CCPC and CCPI, respectively. Our results may be useful for the preparation of precise quantum entangled states and the design of spintronic device.

\section{Model system}
We consider an SO-coupled boson trapped in a driven double well, which is governed by the Hamiltonian
\begin{eqnarray}  \label{eq1}
\hat{H}(t) &=&-\upsilon (t)(\hat{a}_{l}^{\dag}e^{-i\pi \gamma \hat{\sigma} _{y}}
\hat{a}_{r}+H.c.)  \notag\\
&&+\varepsilon (t)\sum_{j }(\hat{n}_{j\uparrow }-\hat{n}_{j\downarrow }).
\end{eqnarray}
Here $\hat{a}_{j}^{\dag}=(\hat{a}_{j\uparrow}^{\dag},\hat{a}_{j\downarrow}^{\dag})$ and $\hat{a}_{j}=(\hat{a}_{j\uparrow },\hat{a}_{j\downarrow })^{T}$ ($T$ denotes the transpose). $\hat{a}_{j\sigma}^{\dag}$ and $\hat{a}_{j\sigma}$ are the creation and annihilation operators of a spin-$\sigma$ ($\sigma=\uparrow, \downarrow$) boson in the $j$th ($j=l,r$) well respectively. $\gamma$ is the effective strength of SO coupling and $\hat{\sigma}_{y}$ is the $y$ component of usual Pauli operator. $H.c.$ represents the Hermitian conjugate of the preceding term. $\hat{n}_{j\sigma}$=$\hat{a}_{j\sigma}^{\dag}\hat{a}_{j\sigma}$ denotes the number operator for spin $\sigma$ in well $j$,
$\upsilon(t)$ and $\varepsilon(t)$ are the time-dependent tunneling amplitude without SO coupling and Zeeman field, respectively. For simplicity, Eq. (1) has been treated as a dimensionless equation in which $\hbar=1$ and the reference frequency $\omega _{0}=0.1 E_r$ ($E_r=k_{L}^2/2M=22.5$kHz is the single-photon recoil energy \cite{lin471}) are set such that the parameters in $\upsilon (t)$ and $\varepsilon (t)$ are in units of $\omega _{0}$, and time $t$ is normalized in units of $\omega _{0}^{-1}$.

Employing the Fock basis $|0,\sigma\rangle(|\sigma,0\rangle)$ to denote the state of a spin-$\sigma$ boson occupying the right (left) well and no atom in the left (right) well, the quantum state of the SO-coupled bosonic system can be expanded as
\begin{equation} \label{eq2}
|\psi (t)\rangle =a_{1}(t)|0, \uparrow \rangle + a_{2}(t)|0, \downarrow \rangle + a_{3}(t)|\uparrow, 0 \rangle + a_{4}(t)|\downarrow, 0\rangle ,
\end{equation}
where $a_{m}(t)(m=1,2,3,4)$ represents the time-dependent probability amplitude of the boson being in state $|0, \sigma \rangle$ or $|\sigma, 0\rangle$, for example, $a_{1}(t)$ represents the time-dependent probability amplitude of the boson being in state $|0, \uparrow \rangle$. The occupation probability reads $P_m(t)=|a_m(t)|^2$, which satisfies the normalization condition $\sum_{m=1}^{4}P_{m}(t)=1$. For the convenience of studying dynamics, we define the population imbalance $Z_{sq}(t)=P_s(t)-P_q(t)$ ($s,q=L,R,1,2,3,4$, $s\neq q$) in which $P_{L}(t)=P_3(t)+P_4(t)$ and $P_{R}(t)=P_1(t)+P_2(t)$ are the total occupation populations of spin-$\sigma$ boson in left and right wells respectively. Inserting Eqs. (1) and (2) into Schr\"{o}dinger equation
$i\frac{\partial |\psi(t) \rangle}{\partial t}=\hat{H}(t)|\psi(t) \rangle$ results in the coupled equations
\begin{eqnarray}  \label{eq3}
i\overset{\cdot }{a}_{1}(t) &=&\varepsilon (t)a_{1}(t)-\upsilon (t)[sin(\pi\gamma )a_{4}(t)+\cos(\pi \gamma )a_{3}(t)],  \notag \\
i\overset{\cdot }{a}_{2}(t) &=&-\varepsilon (t)a_{2}(t)+\upsilon (t)[sin(\pi\gamma )a_{3}(t)-\cos(\pi \gamma )a_{4}(t)],  \notag \\
i\overset{\cdot }{a}_{3}(t) &=&\varepsilon (t)a_{3}(t)+\upsilon (t)[sin(\pi\gamma )a_{2}(t)-\cos(\pi \gamma )a_{1}(t)],  \notag \\
i\overset{\cdot }{a}_{4}(t) &=&-\varepsilon (t)a_{4}(t)-\upsilon (t)[sin(\pi\gamma )a_{1}(t)+\cos(\pi \gamma )a_{2}(t)].\notag \\
\end{eqnarray}

Generally, it is very difficult to get the exact solution of Eq. (3), because of the different forms of the time-dependent modulation functions.
Here, we will attempt to obtain the exact solution of Eq. (3) under the synchronous and asynchronous modulations, respectively.

\section{ANALYTICALLY EXACT SOLUTION UNDER THE SYNCHRONOUS MODULATIONS}

For the synchronous modulations, we mean that the time-dependent driving functions $\varepsilon (t)$ and $\upsilon (t)$ obey the relation $\varepsilon (t)=\beta \upsilon (t)$\cite{hai87}, in which $\beta$ is a proportional constant. In such a case, by introducing a new time variable $\tau=\tau (t)=\int \upsilon (t)dt$, Eq. (3) reduces to
\begin{eqnarray}  \label{eq4}
i\frac{\overset{}{da}_{1}(t)}{d\tau} &=&\beta a_{1}(t)-sin(\pi \gamma)a_{4}(t)-\cos(\pi \gamma)a_{3}(t),  \notag \\
i\frac{\overset{}{da}_{2}(t)}{d\tau} &=&-\beta a_{2}(t)+sin(\pi \gamma)a_{3}(t)-\cos(\pi \gamma)a_{4}(t),  \notag \\
i\frac{\overset{}{da}_{3}(t)}{d\tau} &=&\beta a_{3}(t)+sin(\pi \gamma)a_{2}(t)-\cos(\pi \gamma)a_{1}(t),  \notag \\
i\frac{\overset{}{da}_{4}(t)}{d\tau} &=&-\beta a_{4}(t)-sin(\pi \gamma)a_{1}(t)-\cos(\pi \gamma)a_{2}(t),
\end{eqnarray}
which is the linear differential equations of $\tau$ with constant coefficients. In order to obtain the solutions of Eq. (4), we introduce the stationary solutions, $a_{1}(t)=Ae^{-i\lambda \tau}$, $a_{2}(t)=Be^{-i\lambda \tau}$, $a_{3}(t)=Ce^{-i\lambda \tau}$,
$a_{4}(t)=De^{-i\lambda \tau}$, where $A$, $B$, $C$, and $D$ are constants obeying the normalization condition $|A|^{2}+|B|^{2}+|C|^{2}+|D|^{2}=1$, and $\lambda$ is the characteristic value. Inserting the stationary solutions into Eq. (4), we obtain the four characteristic values $\lambda_{m}$ and the corresponding constants $A_{m}$, $B_{m}$, $C_{m}$, $D_{m}$ for $m=1,2,3,4$ as
\begin{eqnarray}  \label{eq5}
A_{1,2} &=& \frac{1}{\sqrt{2+2\alpha_{\mp}^2}}, B_{1,2} =-D_{1,2}=A_{1,2}\alpha_{\mp}, C_{1,2}=A_{1,2},\notag \\
A_{3,4} &=& \frac{1}{\sqrt{2+2\eta_{\pm}^2}}, B_{3,4} =D_{3,4}=A_{3,4}\eta_{\pm}, C_{3,4}=-A_{3,4},\notag \\
\lambda _{1,2} &=& \mp\sqrt{1+\beta^2-2\beta\cos(\pi \gamma)}, \notag \\
\lambda _{3,4} &=& \mp\sqrt{1+\beta^2+2\beta\cos(\pi \gamma)},
\end{eqnarray}
with constants $\alpha_{\pm}=[-\beta+\cos(\pi \gamma)\pm\sqrt{1+\beta^2-2\beta\cos(\pi \gamma)}]\csc(\pi \gamma)$, $\eta_{\pm}=[\beta+\cos(\pi \gamma)\pm\sqrt{1+\beta^2+2\beta\cos(\pi \gamma)}]\csc(\pi \gamma)$.
Based on Eq. (5), we immediately get the four accurate stationary-like solutions (states) of the system
\begin{eqnarray}  \label{eq6}
|\psi_{m}\rangle &=&(A_{m}\mid 0,\uparrow \rangle
+B_{m}\mid 0,\downarrow \rangle +C_{m}\mid \uparrow ,0\rangle +D_{m}\mid \downarrow ,0\rangle) \nonumber\\ && \times e^{-i\lambda _{m}\tau },
\end{eqnarray}
for $m=1,2,3,4$, which are the exact incoherent destruction of tunneling (IDT) states.

According to the superposition principle of quantum mechanics, we readily obtain the general exact solution of Eq. (4) by employing Eq. (6) to the linear superposition as
\begin{eqnarray}  \label{eq7}
|\psi (t)\rangle=\underset{m=1}{\overset{4}{\sum }}s_{m}|\psi_{m}(t)\rangle.
\end{eqnarray}
Here, $s_m$ for $m=1,2,3,4$ is an arbitrary superposition coefficient determined by the initial condition and normalization. Compared Eq. (7) with Eq. (2), the probability amplitudes are renormalized as
\begin{eqnarray}  \label{eq8}
a_{1}(t)&=&\underset{m=1}{\overset{4}{\sum}}s_{m}A_{m}e^{-i\lambda_{m}\tau}, a_{2}(t)=\underset{m=1}{\overset{4}{\sum}}s_{m}B_{m}e^{-i
\lambda_{m}\tau},\nonumber \\
a_{3}(t)&=&\underset{m=1}{\overset{4}{\sum}}s_{m}C_{m}e^{-i\lambda_{m}\tau}, a_{4}(t)=\underset{m=1}{\overset{4}{\sum}}s_{m}D_{m}e^{-i\lambda_{m}\tau}
\end{eqnarray}
which is the general exact solution of Eq. (3) under the synchronous modulations. Obviously, any of the synchronous combined modulation functions $\upsilon (t)$ and $\varepsilon (t)$ will generate an exact analytical solution to the Schr\"{o}dinger equation, which thus enables us to produce an infinite variety of analytically exact solution (8) for the SO-coupled bosonic system. Here, we will select a square sech-shaped (a pulse-shaped) modulation function as an example to study the precise control of quantum spin dynamics by applying the exact solution under the synchronous modulations for the SO-coupled bosonic double-well system.

Let the modulation function $\upsilon (t)=V$sech$^{2}$$(\Omega t)$,
where $V$ and $\Omega$ are the driving strength and inverse pulse width respectively\cite{hai87},
such that the new time variable $\tau (t)=\frac{V}{\Omega} \tanh (\Omega t)$.
Next, we will mainly discuss the APT with and/or without spin-flipping, the CCPC, and the CCPI between left and right wells based on the general exact solution (8).

\begin{figure*}[htp]\center
\includegraphics[height=1.2in,width=1.8in]{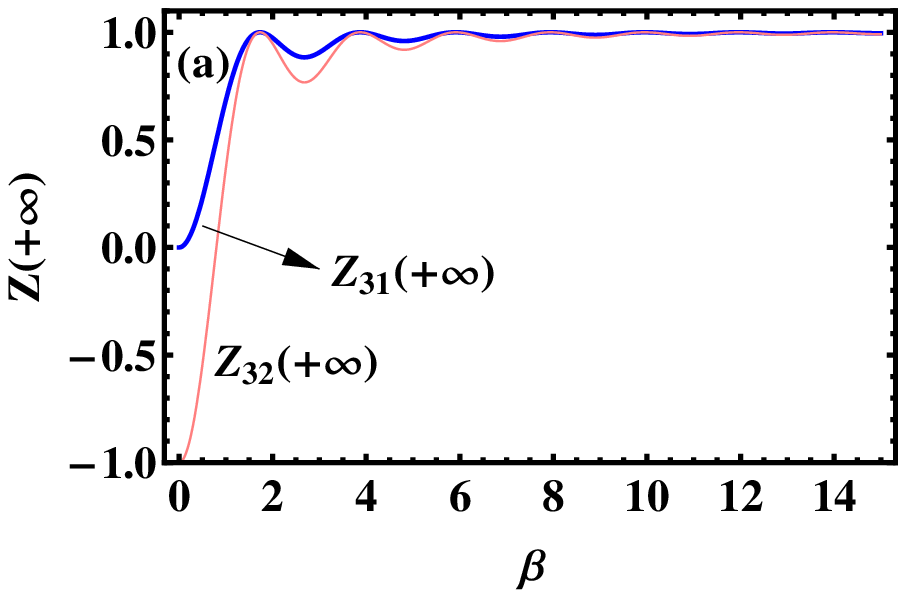}
\includegraphics[height=1.2in,width=1.8in]{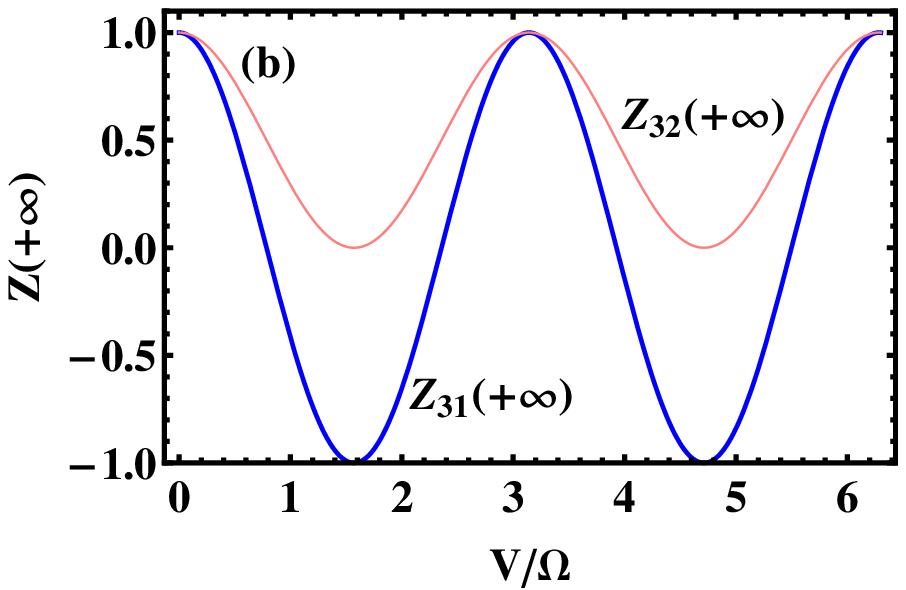}
\includegraphics[height=1.2in,width=1.8in]{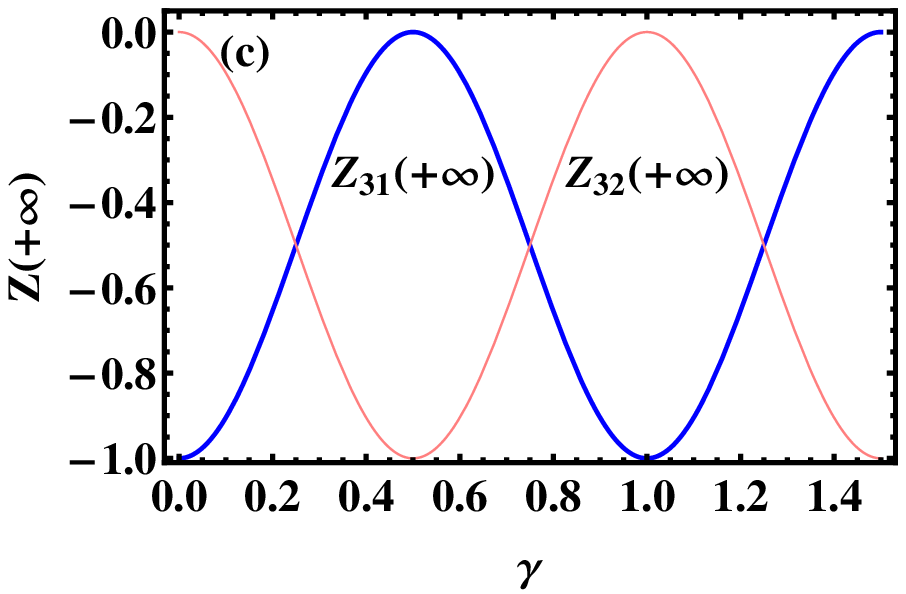}
\includegraphics[height=1.2in,width=1.8in]{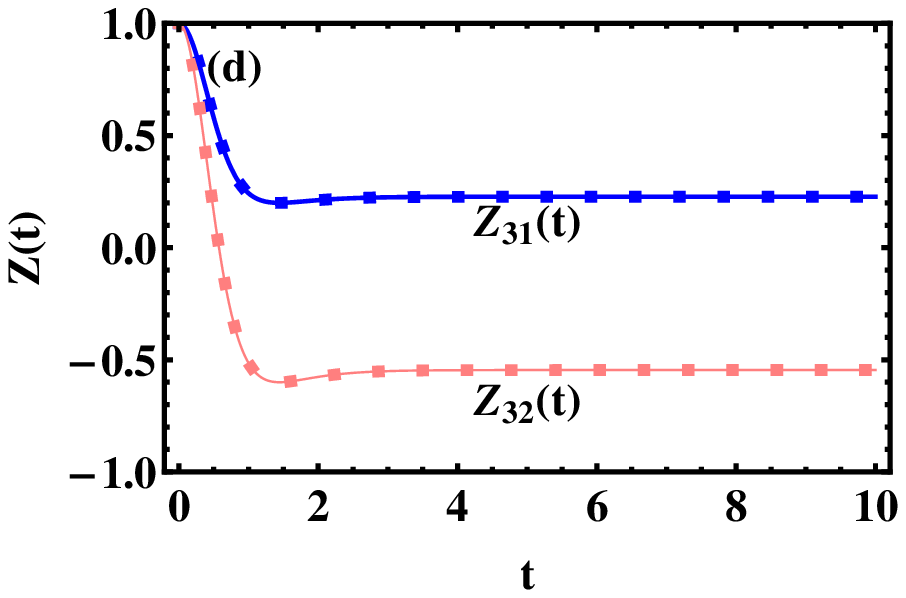}
\includegraphics[height=1.2in,width=1.8in]{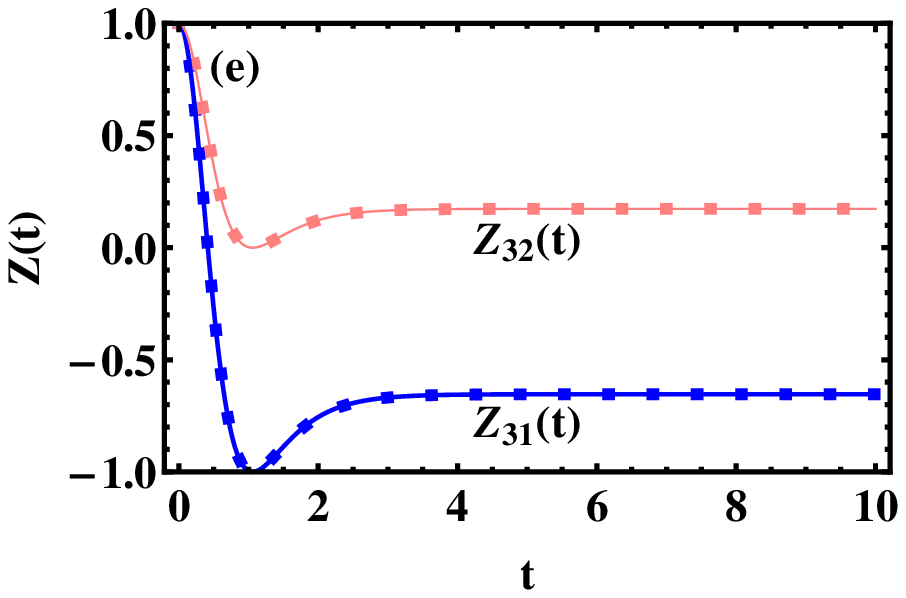}
\includegraphics[height=1.2in,width=1.8in]{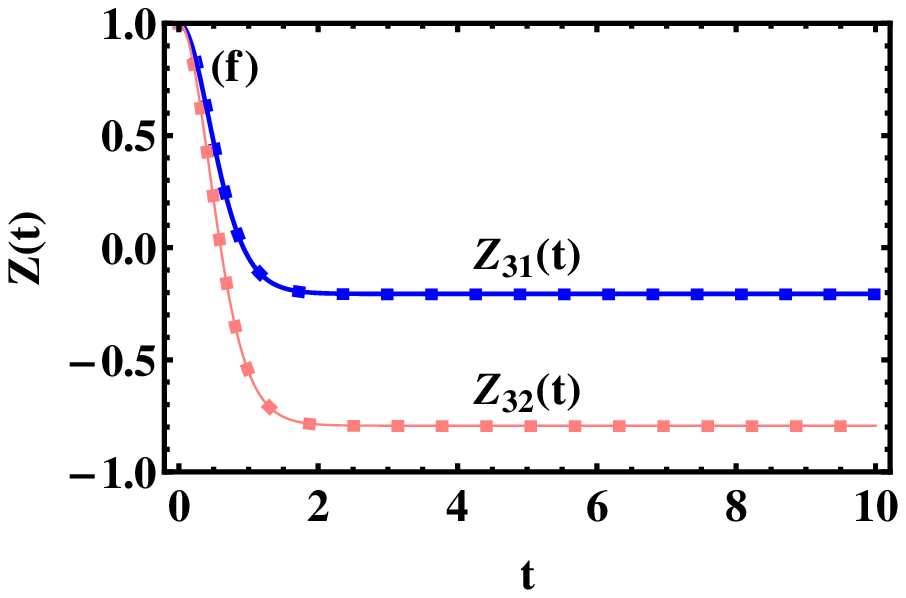}
\caption{\scriptsize {(Color online) (a)-(c) Plots of the asymptotic population imbalances $Z_{31}(+\infty)$ and $Z_{32}(+\infty)$ vs the proportionality constant $\beta$, the ratio of the driving strength and inverse pulse width $\frac{V}{\Omega}$, and the SO coupling strength $\gamma$, respectively. In (a), $\gamma=0.5$, $V=\frac{\pi}{2}$, and $\Omega=1$. In (b), $\gamma=1$ and $\beta=0$. In (c), $\beta=0$, $V=\frac{\pi}{2}$, and $\Omega=1$. (d)-(f) Time evolutions of the population imbalances $Z_{31}(t)$ and $Z_{32}(t)$. The squares label the exact results from Eq. (8), and the solid curves denote the numerical correspondences obtained from Eq. (3). In (d), $\gamma=0.5$, $V=\frac{\pi}{2}$, $\Omega=1$, and $\beta=0.5$. In (e), $\gamma=1$, $\beta=0$ and $\frac{V}{\Omega}=2$. In (f), $\beta=0$, $V=\frac{\pi}{2}$, $\Omega=1$, and $\gamma=0.35$. In (a)-(f), we assume that the initial conditions of this system are $P_3(0)=1$ and $P_m(0)=0$ ($m\neq 3$).}}
\end{figure*}

\begin{figure*}[htp]\center
\includegraphics[height=1.2in,width=1.8in]{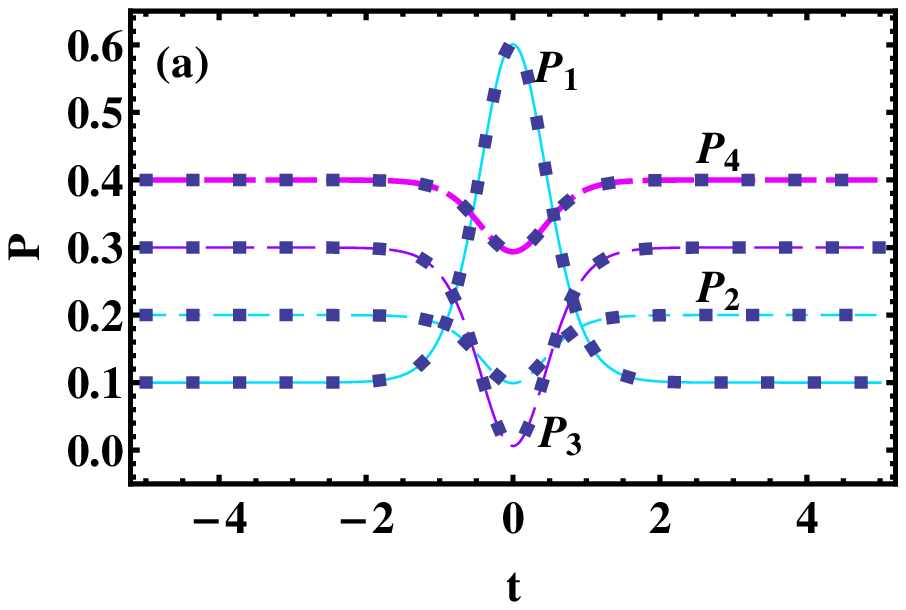}
\includegraphics[height=1.2in,width=1.8in]{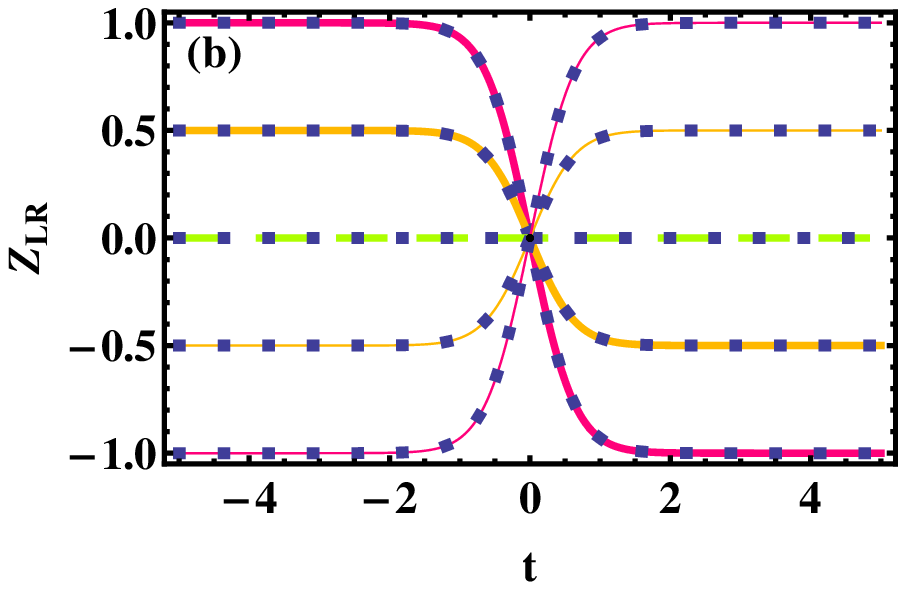}
\includegraphics[height=1.2in,width=1.8in]{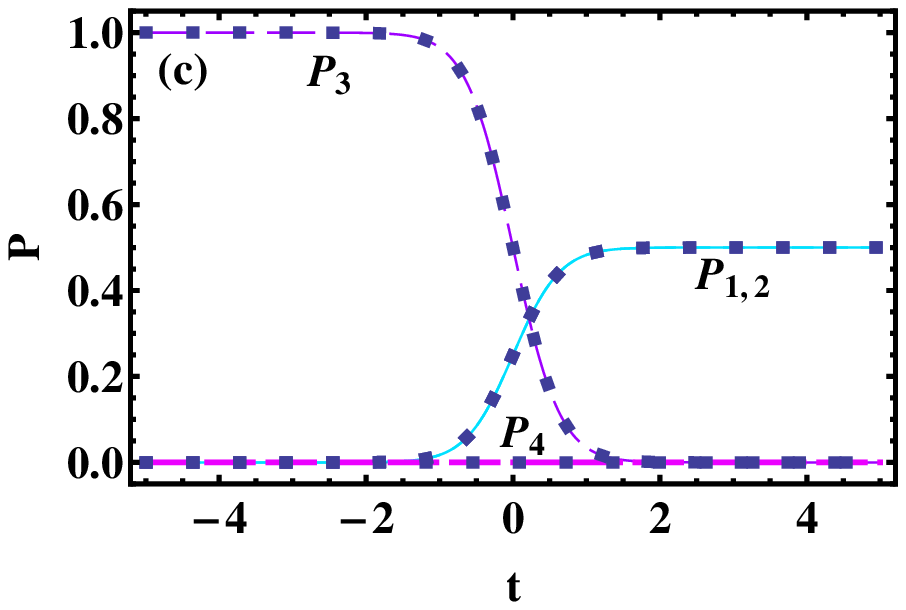}
\caption{{\scriptsize {(Color online) (a)Time evolutions of the probabilities $P_{m}(t)$ for the system parameters $\beta=0$, $V=\frac{\pi}{2}$, $\Omega=1$, $\gamma=0.15$, and the initial conditions $a_{1}(-\infty)=\sqrt{1/10}$, $a_{2}(-\infty)=\sqrt{1/5}$, $a_{3}(-\infty)=\sqrt{3/10}$, and $a_{4}(-\infty)=\sqrt{2/5}$. (b) Time evolutions of the population imbalance $Z_{LR}$ showing the CCPI for the system parameters being the same as those of (a) except for $V=\frac{\pi}{4}$, and the initial conditions $a_{1}(-\infty)=a_{2}(-\infty)=0$, $a_{3}(-\infty)=\sqrt{3/8}$, and $a_{4}(-\infty)=\sqrt{5/8}$ (red thick solid curve); $a_{1}(-\infty)=a_{2}(-\infty)=\sqrt{1/8}$, $a_{3}(-\infty)=1/2$, and $a_{4}(-\infty)=\sqrt{1/2}$ (orange thick solid curve); $a_{1}(-\infty)=a_{2}(-\infty)=1/2$, $a_{3}(-\infty)=\sqrt{1/8}$, and $a_{4}(-\infty)=\sqrt{3/8}$(green dashed curve); $a_{1}(-\infty)=1/2$, $a_{2}(-\infty)=\sqrt{1/2}$, and $a_{3}(-\infty)=a_{4}(-\infty)=\sqrt{1/8}$ (orange thin solid curve); $a_{1}(-\infty)=\sqrt{3/8}$, $a_{2}(-\infty)=\sqrt{5/8}$, and $a_{3}(-\infty)=a_{4}(-\infty)=0$ (red thin solid curve). (c)Time evolutions of the probabilities $P_{m}(t)$ with the initial conditions $a_{3}(-\infty)=1$ and $a_{m}(-\infty)=0(m\neq3)$ for the system parameters being the same as those of (b), except for $\gamma=0.25$. The squares denote the analytical results from Eq. (8), and the curves denote the numerical correspondences obtained from Eq. (3).}}}
\end{figure*}

\subsubsection{APT with and/or without spin-flipping}

From the general exact solution (8), it is easily found that the population imbalance $Z_{sq}(t)$ is a constant at $t\rightarrow +\infty$ and $\tau\rightarrow \frac{V}{\Omega}$.
Assuming the initial state of the system to be state $|\uparrow,0\rangle$ with $P_3(0)=1$ and $P_{m}(0)=0(m\neq3)$, and according to the general exact solution (8), we plot the asymptotic population imbalances $Z_{31}(+\infty)$ and $Z_{32}(+\infty)$ as functions of the proportionality constant $\beta$, the ratio of the driving strength and inverse pulse width $\frac{V}{\Omega}$, and the SO coupling strength $\gamma$ shown in Figs. 1(a)-(c), respectively. In Fig. 1(a), the parameters are taken as $\gamma=0.5$ ($\sin(\pi \gamma)=1$), $V=\frac{\pi}{2}$, and $\Omega=1$. In this case, only quantum tunneling with spin-flipping can occur. Therefore, $Z_{31}(+\infty)$ is equal to $P_{3}(+\infty)$. It is easily seen from Fig. 1(a) that the asymptotic population imbalance $Z_{32}(+\infty)$ can be continuously changed in the regime $Z_{32}(+\infty)\in[-1, 1]$ by adjusting the proportionality constant $\beta$, which means the APT with spin-flipping (or the APT between state $|\uparrow,0\rangle$ and state $|0,\downarrow\rangle$) can be realized. To confirm our result, we select $\beta=0.5$ in Fig. 1(a) as an example to plot the time evolutions of the population imbalances (see Fig. 1(d)). From Fig. 1(d), one can get $Z_{32}(+\infty)\approx-0.5456$ and $Z_{31}(+\infty)\approx0.2272$ which are in agreement with the results shown in Fig. 1(a). Furthermore, we find that as the proportionality constant $\beta$ increases, the asymptotic population imbalances $Z_{31}(+\infty)$ and $Z_{32}(+\infty)$ will tend to one in Fig. 1(a), which mean the coherent destruction of tunneling (CDT) happens (not shown here).

In Fig. 1(b), we set the parameters $\beta=0$ and $\gamma=1$ ($\sin(\pi \gamma)=0$). Such that only quantum tunneling without spin-flipping can be performed and $Z_{32}(+\infty)=P_{3}(+\infty)$. The asymptotic population imbalance $Z_{31}(+\infty)$ can be continuously adjusted in the regime $Z_{31}(+\infty)\in[-1, 1]$ via regulating the ratio value $\frac{V}{\Omega}$, which means the APT without spin-flipping (or the APT between state $|\uparrow,0\rangle$ and state $|0,\uparrow\rangle$) can be actualized. For example, we fix $\frac{V}{\Omega}=2$ in Fig. 1(b) to plot the time evolutions of the population imbalances (see Fig. 1(e)). It can be seen from Fig. 1(e) that the asymptotic population imbalances $Z_{32}(+\infty)\approx0.1732$ and $Z_{31}(+\infty)\approx-0.6536$ which are consistent with the results shown in Fig. 1(b). Specially, when $\frac{V}{\Omega}=(n+\frac{1}{2})\pi (n=0,1,2,...)$ in Fig. 1(b), the asymptotic population imbalance $Z_{31}(+\infty)=-1$ which means the completely tunneling without spin-flipping from the initial state $|\uparrow,0\rangle$ to the final state $|0,\uparrow\rangle$ takes place (not shown here).

In Fig. 1(c), the parameters $\beta=0$, $V=\frac{\pi}{2}$, and $\Omega=1$ are fixed. According to Eq. (8) and normalization condition, one can get $P_{3}(+\infty)=P_{4}(+\infty)=0$, $P_{1}(+\infty)+P_{2}(+\infty)=1$, and $Z_{31}(+\infty)+Z_{32}(+\infty)=-1$. It is obviously seen from Fig. 1(c) that the asymptotic population imbalances $Z_{31}(+\infty)$ and $Z_{32}(+\infty)$ oscillate periodically with the SO coupling strength $\gamma$ in the region $Z_{31}(+\infty), Z_{32}(+\infty)\in[0,-1]$, which mean the tunneling probabilities with and without spin-flipping can be controlled, namely, the APT with and without spin-flipping can be manipulated. As an example, we take $\gamma=0.35$ in Fig. 1(c) to plot the time evolutions of the population imbalances (see Fig. 1(f)). In Fig. 1(f), the asymptotic population imbalances $Z_{31}(+\infty)\approx-0.2061$ and $Z_{32}(+\infty)\approx-0.7939$ that are accordant with the results shown in Fig. 1(c). In particular, $Z_{31}(+\infty)=Z_{32}(+\infty)=-0.5$ when $\gamma=\frac{1}{4}(2n+1)(n=0,1,2,...)$, which means the spin-flipping and spin-conserving quantum tunneling with equal probability will simultaneously happen (not shown here).

\subsubsection{CCPC and CCPI between left and right wells}

Now, we mainly discuss the CCPC and CCPI between left and right wells. From the exact solution (8), we surprisingly find that when the parameters satisfy
\begin{eqnarray}  \label{eq9}
\beta=0,  \ \frac{2V}{\Omega}=n \pi, \ n=1,2,3,... ,
\end{eqnarray}
the probability $P_{m}(-\infty)=P_{m}(+\infty)$ ($m=1,2,3,4$) which means the final occupation probability of particle returns to the initial occupation probability as time goes from  $-\infty$ to $+\infty$ and the final state is controllable. Thus, we call it \emph{CCPC}. Here, we set the initial conditions $a_{1}(-\infty)=\sqrt{1/10}$, $a_{2}(-\infty)=\sqrt{1/5}$, $a_{3}(-\infty)=\sqrt{3/10}$, $a_{4}(-\infty)=\sqrt{2/5}$, and take the parameters $\beta=0$, $V=\frac{\pi}{2}$, $\Omega=1$, and $\gamma=0.15$ to plot the time evolutions of the probabilities as shown in Fig. 2(a). From Fig. 2(a), one can find the probability $P_{m}(-\infty)=P_{m}(+\infty)$, that is to say, the CCPC occurs. But, it is worth noting that the curves have oscillation around t=0, which imply the quantum transitions of particle between Fock states.

Further, from the accurate solution (8) we also notice that when the parameters satisfy
\begin{eqnarray}  \label{eq10}
\beta=0, \  \frac{2V}{\Omega}=(n+\frac{1}{2})\pi, \  n=0,1,2,... ,
\end{eqnarray}
the population imbalances $Z_{LR}(-\infty)=-Z_{LR}(+\infty)$ which mean the total populations in left well and that in right well inverse as time goes from $-\infty$ to $+\infty$ and the final populations in left and right wells can be controlled by setting appropriately initial conditions. So we call it \emph{CCPI} between left and right wells. As an example, we take the same parameters as those of Fig. 2(a) except for $V=\frac{\pi}{4}$ and plot the time evolutions of the population imbalance $Z_{LR}(t)$ for different initial conditions as shown in Fig. 2(b). It is obviously seen that $Z_{LR}(-\infty)=-Z_{LR}(+\infty)$. Therefore, the CCPI between left and right wells happens. That is to say, the CCPI still exists under the synchronous modulations in the SO-coupled bosonic system (four-level system), which generalizes the result obtained in the two-level system under asynchronous modulation (see Ref. \cite{hai87}).
Furthermore, what's particularly interesting is that we find the number of oscillating peaks (valleys) of the probability $P_{m}(t)$ and the population imbalance $Z_{LR}(t)$ near $t=0$ in Figs. 2(a) and (b) are equal to the corresponding value of $n$ (e.g., $n=1$ peak for the curve of the probability $P_1$ near $t=0$ in Fig. 2(a)), except for the case of the initial population imbalance $Z_{LR}(-\infty)=0$ in Fig. 2(b). Not only that we also find when the initial condition is $P_{3}(-\infty)=1$ and the SO coupling strength $\gamma=n+0.25$ ($n=0,1,2,...$) for the case of CCPI between left and right wells, the probabilities $P_{1}(+\infty)=P_{2}(+\infty)=0.5$ which mean the spin-flipping and spin-conserving tunneling with equal probability occurs simultaneously as shown in Fig. 2(c).

\section{ANALYTICALLY EXACT SOLUTIONS UNDER THE ASYNCHRONOUS MODULATIONS}

In this section, by using the tanh-shaped modulation $\varepsilon(t)=\varepsilon \tanh(\chi t)$ and sech-shaped modulation $\upsilon(t)$=$\upsilon$ sech$(\chi t)$, our aim is to get the exact solutions of the asynchronous modulation system for the cases of spin-conserving and spin-flipping tunneling respectively. Then, applying those exact solutions to study the CCPC and CCPI with or without spin-flipping. From Eq. (3), it can be seen that when $\sin(\pi\gamma)=0$, $a_{1}$ is only coupled with $a_{3}$ and $a_{2}$ is only coupled with $a_{4}$, which mean only the quantum tunneling without spin-flipping can occur. In order to investigate the exact CCPC and CCPI without spin-flipping, without loss of generality, we set $\cos(\pi\gamma)=1$ (e.g., $\gamma=2$), such that Eq. (3) reduces to two set of decoupled subequations as
\begin{subequations}  \label{XX}
\begin{align}
i\overset{\cdot }{a}_{1}(t)&=\varepsilon (t)a_{1}(t)-\upsilon (t)a_{3}(t),  \notag \\
i\overset{\cdot }{a}_{3}(t)&=\varepsilon (t)a_{3}(t)-\upsilon (t)a_{1}(t), \label{Za}\\
i\overset{\cdot }{a}_{2}(t)&=-\varepsilon (t)a_{2}(t)-\upsilon (t)a_{4}(t),  \notag \\
i\overset{\cdot }{a}_{4}(t)&=-\varepsilon (t)a_{4}(t)-\upsilon (t)a_{2}(t).\label{Zb}
\end{align}
\end{subequations}

Here, we solve Eq. (11a) as an example. By introducing the transformation $a_1(t)=b_1(t) e^{-i \int \varepsilon (t) dt}$ and $a_3(t)=b_3(t) e^{-i \int \varepsilon (t) dt}$ and inserting them to Eq. (11a), such that the Eq. (11a) reduces to
\begin{eqnarray}  \label{eq12}
i\overset{\cdot }{b}_{1}(t)&=-\upsilon (t)b_{3}(t),  \notag \\
i\overset{\cdot }{b}_{3}(t)&=-\upsilon (t)b_{1}(t).
\end{eqnarray}

By using a time scale, $\tau=\tau (t)=\int \upsilon (t)dt$, and utilizing the relation $\frac{db_1(t)}{dt}=\frac{db_1(t)}{d\tau}\frac{d\tau}{dt}=\frac{db_1(t)}{d\tau}\upsilon (t)$,
the Eq. (12) can be simplified to the form
\begin{eqnarray}  \label{eq13}
i\frac{db_1(t)}{d\tau}&=-b_{3}(t),  \notag \\
i\frac{db_3(t)}{d\tau}&=-b_{1}(t).
\end{eqnarray}

Obviously, the general solution of Eq. (13) can be easily obtained as
\begin{eqnarray}  \label{eq14}
b_1(t)&=A_{+}e^{i \tau}+A_{-}e^{-i \tau},  \notag \\
b_3(t)&=A_{+}e^{i \tau}-A_{-}e^{-i \tau}.
\end{eqnarray}
Such that we can get the exact solution of Eq. (11a) as
\begin{eqnarray}  \label{eq15}
a_1(t)&=A_{+}e^{i \int [\upsilon (t)-\varepsilon (t)]dt}+A_{-}e^{-i \int [\upsilon (t)+\varepsilon (t)]dt},  \notag \\
a_3(t)&=A_{+}e^{i \int [\upsilon (t)-\varepsilon (t)]dt}-A_{-}e^{-i \int [\upsilon (t)+\varepsilon (t)]dt}.
\end{eqnarray}

Similarly, we can obtain the precise solution of Eq. (11b) as
\begin{eqnarray}  \label{eq16}
a_2(t)&=B_{+}e^{i \int [\upsilon (t)+\varepsilon (t)]dt}+B_{-}e^{-i \int [\upsilon (t)-\varepsilon (t)]dt},  \notag \\
a_4(t)&=B_{+}e^{i \int [\upsilon (t)+\varepsilon (t)]dt}-B_{-}e^{-i \int [\upsilon (t)-\varepsilon (t)]dt}.
\end{eqnarray}
Here, the complex superposition constants $A_{\pm}$ and $B_{\pm}$ are determined by the initial conditions and normalization respectively.

These solutions, Eqs. (15) and (16), are the exact ones corresponding for the case of spin-conserving tunneling
for arbitrary modulation functions $\upsilon (t)$ and $\varepsilon (t)$. Here, we only consider the asynchronous modulations
$\varepsilon(t)=\varepsilon\tanh(\chi t)$ and $\upsilon(t)$=$\upsilon$ sech$(\chi t)$. In such case, from Eq.(15), we surprisingly find when the parameters satisfy
\begin{eqnarray}  \label{eq17}
\sin(\pi \upsilon/\chi)=0,
\end{eqnarray}
the populations $P_3(-\infty)=P_3(+\infty)$ and $P_1(-\infty)=P_1(+\infty)$ with the population
imbalances $Z_{31}(-\infty)=Z_{31}(+\infty)=\pm 2(A_{+}A_{-}^{\ast}+A_{+}^{\ast}A_{-})$ (here, $\pm$ sign corresponds to $\cos(\pi \upsilon/\chi)=\mp 1$ respectively), which mean the CCPC occurs as time goes from $-\infty$ to $+\infty$. As an example, we take the parameters $\gamma=2$, $\varepsilon=1$, $\upsilon=1$, $\chi=1$ and set the different initial conditions to plot the time evolutions of the population imbalance $Z_{31}(t)$ in Fig. 3 (a). It can be seen that for different initial conditions the population imbalances remain unchanged as $Z_{31}(-\infty)=Z_{31}(+\infty)$, namely, the CCPC happens.

When the parameters satisfy
\begin{eqnarray}  \label{eq18}
\cos(\pi \upsilon/\chi)=0,
\end{eqnarray}
the populations $P_3(-\infty)=P_1(+\infty)$ and $P_1(-\infty)=P_3(+\infty)$ with the population imbalances $Z_{31}(-\infty)=-Z_{31}(+\infty)=\pm 2 i(A_{+}A_{-}^{\ast}-A_{+}^{\ast}A_{-})$ (here, $\pm$ sign corresponds to $\sin(\pi \upsilon/\chi)=\pm 1$ respectively), which mean the occurrence of CCPI without spin-flipping as time $t=-\infty\rightarrow +\infty$. As an example, we take the same parameters and initial conditions as those of Fig. 3(a), except for $\chi=2$, and plot the time evolutions of the population imbalance $Z_{31}(t)$ in Fig. 3 (b). It can be seen that the population imbalance $Z_{31}(-\infty)=-Z_{31}(+\infty)$ under the different initial conditions, which means the CCPI occurs.

It is worth noting that the time-evolution curves of the population imbalance $Z_{31}(t)$ have oscillation around $t=0$ and the number of oscillating peak (or valley) for each curve is equal to the value $\upsilon/\chi$ and $\upsilon/\chi-1/2$ in Figs. 3(a) and (b) respectively, e.g., $\upsilon/\chi=1$ in Fig. 3(a) and $\upsilon/\chi-1/2=0$ in Fig. 3(b). Specially, when the initial population imbalance $Z_{31}(-\infty)=0$, the population of particle will keep standstill which means the occurrence of CDT, see Figs. 3(a) and (b). The analytical results (the squares) from Eq. (15) are in complete agreement with the numerical results (the curves) from Eq. (11a). For the exact solution (16), we can obtain the similar results.

\begin{figure}[tph]\center
\includegraphics[height=0.9in,width=1.4in]{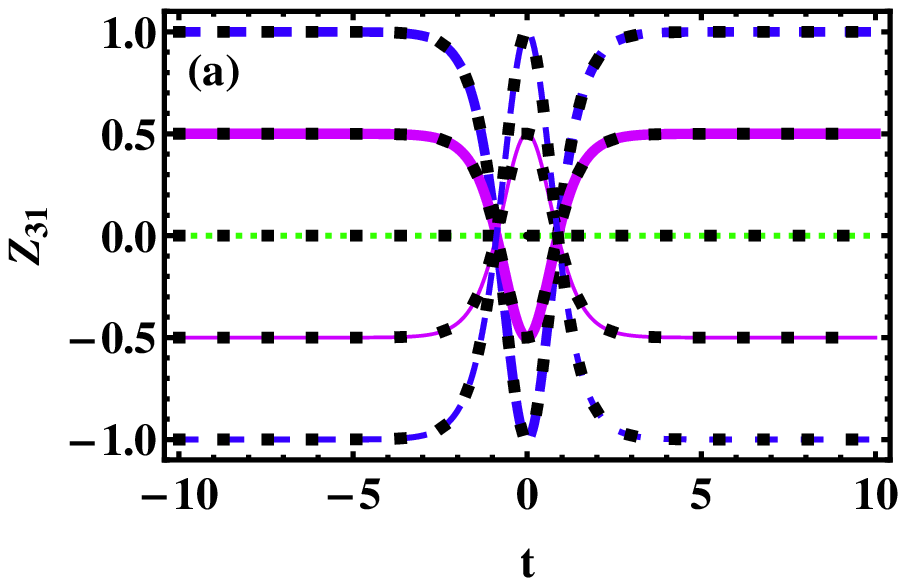}
\includegraphics[height=0.9in,width=1.4in]{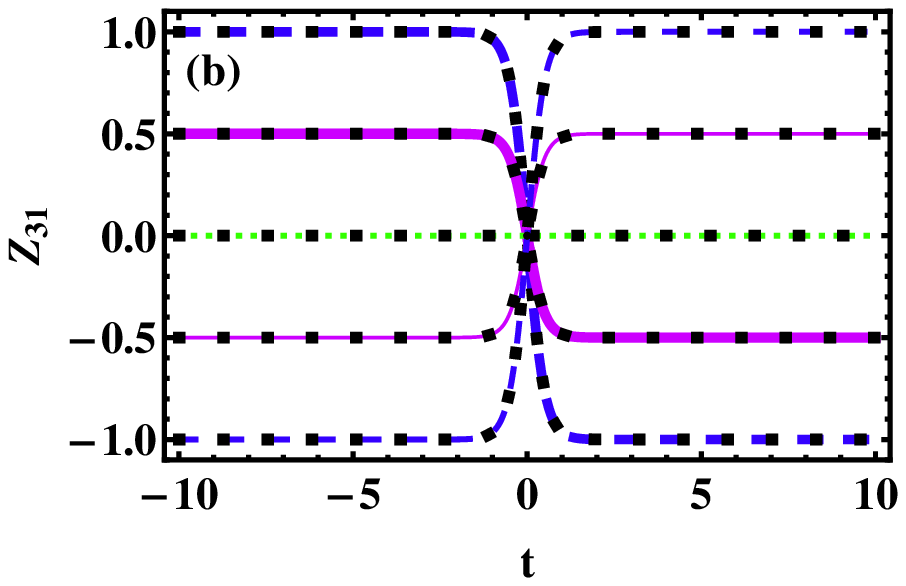}
\includegraphics[height=1.0in,width=1.45in]{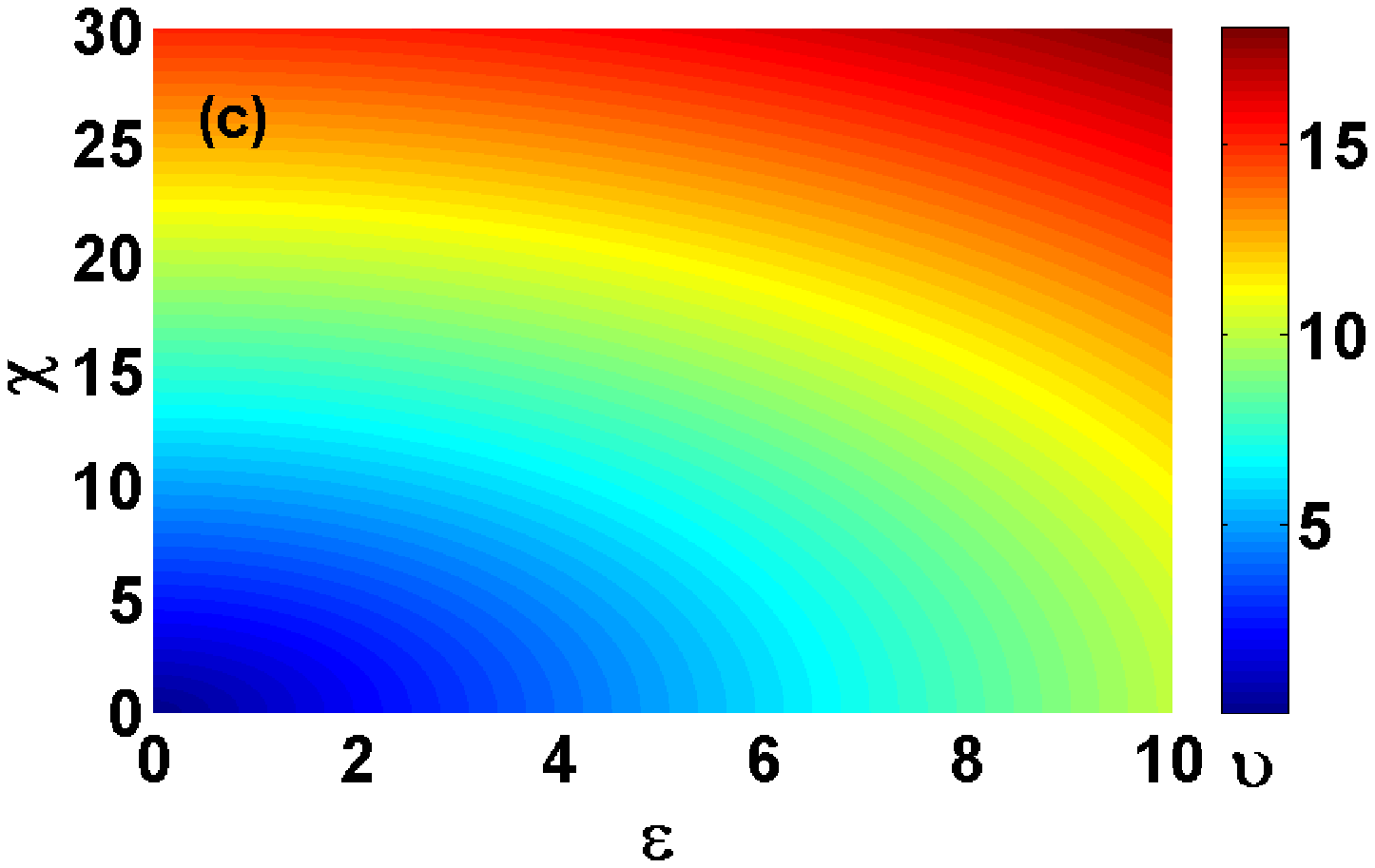}
\includegraphics[height=0.9in,width=1.4in]{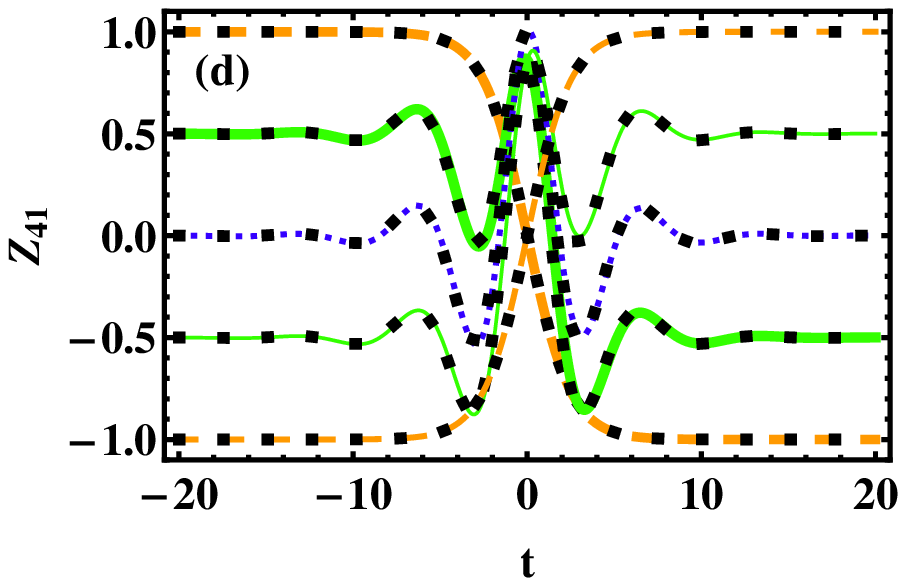}
\caption{{\scriptsize {(Color online) (a) Time evolutions of the population imbalance $Z_{31}$ showing the CCPC for the system parameters $\gamma=2$, $\varepsilon=1$, $\upsilon=1$, $\chi=1$, and the initial conditions $a_{3}(-\infty)=1$ and $a_{m}(-\infty)=0$ ($m\neq3$) (blue thick dashed curve); $a_{1}(-\infty)=\frac{1}{2}$, $a_{3}(-\infty)=\frac{\sqrt{3}}{2}$, and $a_{m}(-\infty)=0$ ($m\neq1,3$) (purple thick solid curve); $a_{1}(-\infty)=a_{3}(-\infty)=\frac{1}{\sqrt{2}}$ and $a_{m}(-\infty)=0$ ($m\neq1,3$)(green dotted curve); $a_{1}(-\infty)=\frac{\sqrt{3}}{2}$, $a_{3}(-\infty)=\frac{1}{2}$, and $a_{m}(-\infty)=0$ ($m\neq1,3$)(purple thin solid curve); $a_{1}(-\infty)=1$ and $a_{m}(-\infty)=0$ ($m\neq1$)(blue thin dashed curve).(b) Time evolutions of the population imbalance $Z_{31}$ showing the CCPI without spin-flipping for the system parameters and initial conditions being the same as those of (a), except for $\chi=2$.(c) Plot of the parameter $\upsilon$ as a function of $\chi$ and $\varepsilon$. (d) Time evolutions of the population imbalance $Z_{41}$ showing the CCPI with spin-flipping for the system parameters $\gamma=0.5$, $\varepsilon=\sqrt{0.21}$, $\upsilon=0.5$, $\chi=0.4$, and the initial conditions $a_{4}(-\infty)=1$ and $a_{m}(-\infty)=0$ ($m\neq4$) (orange thick dashed curve); $a_{4}(-\infty)=\frac{\sqrt{3}}{2}$, $a_{1}(-\infty)=\frac{1}{2}$, and $a_{m}(-\infty)=0$ ($m\neq1,4$) (green thick solid curve); $a_{4}(-\infty)=a_{1}(-\infty)=\frac{1}{\sqrt{2}}$ and $a_{m}(-\infty)=0$ ($m\neq1,4$) (blue dotted curve); $a_{4}(-\infty)=\frac{1}{2}$, $a_{1}(-\infty)=\frac{\sqrt{3}}{2}$, and $a_{m}(-\infty)=0$ ($m\neq1,4$)(green thin solid curve); $a_{1}(-\infty)=1$ and $a_{m}(-\infty)=0$ ($m\neq1$) (orange thin dashed curve).}}}
\end{figure}

Next, we study the exact CCPI with spin-flipping and set $\cos(\pi\gamma)=0$ in Eq. (3) which means only quantum tunneling with spin-flipping can occur. Without loss of generally, we fix $\sin(\pi\gamma)=1$ (e.g., $\gamma=0.5$), such that Eq. (3) reduces to two set of decoupled subequations as
\begin{subequations}\label{XX}
\begin{align}
i\overset{\cdot}{a}_{1}(t)&=\varepsilon(t)a_{1}(t)-\upsilon (t)a_{4}(t), \notag \\
i\overset{\cdot}{a}_{4}(t)&=-\varepsilon(t)a_{4}(t)-\upsilon (t)a_{1}(t),\label{Za}\\
i\overset{\cdot}{a}_{2}(t)&=-\varepsilon (t)a_{2}(t)+\upsilon (t)a_{3}(t),  \notag \\
i\overset{\cdot}{a}_{3}(t)&=\varepsilon(t)a_{3}(t)+\upsilon (t)a_{2}(t). \label{Zb}
\end{align}
\end{subequations}

Note that the Eqs. (19a) and (19b) are analogous to Eq. (3) in Ref. \cite{hai87}, so we use the established method in Ref. \cite{hai87} to solve Eq. (19a) as an example to show the exact CCPI with spin-flipping. Let
$a_4(t)=\sqrt{sech(\chi t)}a(t)$ and from Eq. (19a) we can obtain the second-order equation
\begin{eqnarray}  \label{eq20}
&&\ddot{a}(t)+[\upsilon^{2} - \frac{1}{2}\chi^{2}-i\varepsilon\chi + (\frac{1}{4}\chi^{2}+\varepsilon^{2}-\upsilon^{2})  \notag \\
&&\times\tanh(\chi t)^{2}] a(t)=0,
\end{eqnarray}
which is exactly solvable by the use of hypergeometric functions\cite{hai78}. Here, in order to get a simple solution, we select the driving parameters to satisfy
\begin{eqnarray}  \label{eq21}
\frac{1}{4}\chi^{2}+\varepsilon^{2}-\upsilon^{2}=0,
\end{eqnarray}
which is shown in Fig. 3(c). Based on Eq. (21), we can obtain the simple general solution of Eq. (20) as $a(t)=C_{+}e^{(i\varepsilon+\frac{1}{2}\chi)t}+C_{-}e^{-(i\varepsilon+\frac{1}{2}\chi)t}$ with constant $C_{\pm}$ determined by the initial condition. Such that the simple exact solution of Eq. (19a) is obtained as
\begin{eqnarray}  \label{eq22}
a_{1}(t)&=&-\frac{i(\chi+2i\varepsilon)\sqrt{sech(\chi t)}}{2\upsilon} \notag \\
&&\times [C_{+}e^{-\frac{1}{2}(\chi-2i\varepsilon)t}-C_{-}e^{\frac{1}{2}(\chi-2i\varepsilon)t}], \notag \\
a_{4}(t)&=&\sqrt{sech(\chi t)}[C_{+}e^{(i\varepsilon+\frac{1}{2}\chi)t}+C_{-}e^{-(i\varepsilon+\frac{1}{2}\chi)t}]. \notag \\
\end{eqnarray}
It is worth noting that Eq. (22) implies $P_{1}(-\infty)=P_{4}(+\infty)=2|C_{+}|^{2}$ and $P_{4}(-\infty)=P_{1}(+\infty)=2|C_{-}|^{2}$ with the population imbalance $Z_{41}(-\infty)=-Z_{41}(+\infty)$, which mean the exact CCPI with spin-flipping can be implemented.
As an example, we take the SO coupling strength $\gamma=0.5$ and fix the parameters $\varepsilon=\sqrt{0.21}$, $\upsilon=0.5$ and $\chi=0.4$ obeying Eq. (21) to plot the time evolutions of the population imbalance $Z_{41}(t)$ for the different initial conditions as shown in Fig. 3(d). It is obviously seen that the population imbalances $Z_{41}(-\infty)=-Z_{41}(+\infty)$, which mean the CCPI with spin-flipping happens. Here, we note that the time-evolution curve for the initial population imbalance $Z_{41}(-\infty)=0$ has oscillation, which is different from the case of CCPI without spin-flipping (see Fig. 3(b)). The exact results (the squares) from Eq. (22) are in good agreement with the numerical results (the curves) from Eq. (19a).

By using the homologous method of solving Eq. (19a), we readily get the simple exact solution of Eq. (19b) as
\begin{eqnarray} \label{eq23}
a_{2}(t)&=&- \frac{i(\chi-2i\varepsilon)}{2\upsilon\sqrt{sech(\chi t)}}
 [D_{+}(\tanh(\chi t)+1)e^{-\frac{1}{2}(\chi-2i\varepsilon)t}\notag \\
 &&+D_{-}(\tanh(\chi t)-1)e^{\frac{1}{2}(\chi-2i\varepsilon)t}], \notag \\
a_{3}(t)&=&\sqrt{sech(\chi t)}[D_{+}e^{(i\varepsilon-\frac{1}{2}\chi)t}+ D_{-}e^{-(i\varepsilon-\frac{1}{2}\chi)t}],
\end{eqnarray}
where the constant $D_{\pm}$ is determined by the initial condition and the driving parameters must also obey Eq. (21). Likewise, it is also noticed in Eq. (23) that $P_{3}(-\infty)=P_{2}(+\infty)=2|D_{+}|^{2}$ and $P_{2}(-\infty)=P_{3}(+\infty)=2|D_{-}|^{2}$ with the population imbalance $Z_{32}(-\infty)=-Z_{32}(+\infty)$(not shown here). Note that the above results obtained in the case of spin-flipping are similar to ones obtained
in the two-level system without SO coupling (see Ref. \cite{hai87}), which goes against common intuition.

\section{CONCLUSION and DISCUSSION}

In summary, we have proposed a simple method of combined modulations to obtain the analytical exact solutions for an SO-coupled boson trapped in a driven double-well potential. For the cases of synchronous combined modulations
and the non-spin-flipping tunneling, we have obtained the general analytical precise solutions of the system respectively. For the spin-flipping tunneling case under asynchronous combined modulations, we have gotten the particular accurate solutions in simple form when the driving parameters are appropriately chosen. Based on these obtained accurate solutions, we have revealed some intriguing spin dynamical phenomena. Under synchronous combined modulations, the APT with and/or without spin-flipping, the CCPC, and the CCPI between left and right wells have been displayed. Under asynchronous combined modulations, we have also performed the CCPC and the CCPI with spin-conserving or with spin-flipping, respectively. Surprisingly, we have found that the CCPI exists in the cases of synchronous and asynchronous combined modulations in the SO-coupled cold atomic system (four-level system), which generalizes the result obtained in the case of asynchronous modulation in the two-level system without SO coupling (see Ref. \cite{hai87}). The results may have potential applications in the qubit manipulation and the preparation of accurate quantum entangled states.

\section*{ACKNOWLEDGMENTS}
This work was supported by the Hunan Provincial Natural Science Foundation of China under Grants No. 2021JJ30435 and No. 2017JJ3208, the Scientific Research Foundation of Hunan Provincial Education Department under Grants No. 21B0063 and No. 18C0027, and the National Natural Science Foundation of China under Grant No. 11747034.

\end{document}